\documentclass[graphics,floatfix, nofootinbib,
ightenlines,nobibnotes,aps,12pt,numberedappendix,appendixfloats]{revtex4-1}
\usepackage{epsfig}
\usepackage{amssymb,amsmath}
\usepackage{amsfonts}
\usepackage{amssymb}
\usepackage{graphicx}
\usepackage{breqn}
\usepackage{caption}
\usepackage{subfigure}
\usepackage{epstopdf}
\usepackage{array}
\usepackage{dcolumn}   
\usepackage{bm}  
\usepackage{comment}
\usepackage{etoolbox}
\usepackage{fancyhdr}
\newcommand{\bea}{\begin{eqnarray}}
\newcommand{\eea}{\end{eqnarray}}

 \makeatletter
\def\alt{\mathrel{\mathpalette\gl@align<}}
\def\agt{\mathrel{\mathpalette\gl@align>}}
\def\gl@align#1#2{\lower.6ex\vbox{\baselineskip\z@skip\lineskip\z@
\ialign{$\m@th#1\hfil##\hfil$\crcr#2\crcr\sim\crcr}}} \makeatother

\begin{document}
%
\vspace*{1.0cm}

\begin{center}
\baselineskip 20pt 
{\Large\bf 
Radiative Breaking of the Minimal Supersymmetric Left-Right Model
}
\vspace{1cm}

{\large 
Nobuchika Okada~\footnote{okadan@ua.edu}  
and  Nathan Papapietro~\footnote{npapapietro@crimson.ua.edu}
}
\vspace{.5cm}

{\baselineskip 20pt \it
Department of Physics and Astronomy, University of Alabama, Tuscaloosa, AL35487, USA} 

\vspace{.5cm}

\vspace{1.5cm} {\bf Abstract}
\end{center}

We study a variation to the SUSY Left-Right symmetric model based on the gauge group 
   $SU(3)_c\times SU(2)_L\times SU(2)_R\times U(1)_{BL}$. 
Beyond the quark and lepton superfields we only introduce a second Higgs bidoublet 
  to produce realistic fermion mass matrices. 
This model does not include any $SU(2)_R$ triplets. 
We calculate renormalization group evolutions of soft SUSY parameters at the one-loop level down to low energy. 
We find that an $SU(2)_R$ slepton doublet acquires a negative mass squared at low energies, 
  so that the breaking of $SU(2)_R\times U(1)_{BL}\rightarrow U(1)_Y$ is realized 
  by a non-zero vacuum expectation value of a right-handed sneutrino. 
Small neutrino masses are produced through neutrino mixings with gauginos. 
Mass limits on the $SU(2)_R\times U(1)_{BL}$ sector are obtained by direct search results at the LHC 
 as well as lepton-gaugino mixing bounds from the LEP precision data.

\thispagestyle{empty}

\newpage

\addtocounter{page}{-1}

\baselineskip 18pt

\section{Introduction} 
Nature at low energies can be described by a vector-like model known as Quantum Electrodynamics (QED).  
Adding the strong interactions into the mix, nature retains its indifference to a fields' handedness.  
At higher energies, we encounter the Standard Model (SM) which is a chiral theory 
   that is broken down into QED via Electroweak Symmetry Breaking (EWSB). 
Among the fermions in the SM only left-handed fields interact under $SU(2)_L$.  
This question of why does such a parity violation exist as well many others are not cannot be answered by the SM alone. 
Motivation for nature returning to vector-like at TeV scales and higher has led to Left-Right symmetric Models (LRMs) being introduced.  
The first LRM was a broken Pati-Salam model~\cite{PhysRevD.10.275} introduced in \cite{PhysRevD.11.566} 
  with the gauge group $SU(3)_c\times SU(2)_L\times SU(2)_R\times U(1)_{BL}$. 
The LR symmetry must be broken at low energies, TeV scale LRMs are being once again considered 
  from the view point of the Large Hadron Collider (LHC) experiments. 
The current lower bound on the $SU(2)_R$ charged gauge boson ($W_R$) is found to be around 3 TeV~\cite{wrexp} 
  (see also \cite{PhysRevD.82.055022} on the lower bound from rare decay processes).

Historically the first type of LR symmetry breaking was done by a $SU(2)_R$ doublet Higgs field\cite{PhysRevD.11.2558,PhysRevD.12.1502}. After the introduction of the seesaw mechanism~\cite{Minkowski1977421}, breaking LR symmetry 
  by $SU(2)_L$ and $SU(2)_R$ triplets was considered. 
This case has new sets of unnaturalness problems with keeping the $SU(2)_L$ triplet vacuum expectation value (VEV)  
   at the neutrino mass scale~\cite{PhysRevD.44.837}. 
Its minimal superymmetric (SUSY) extensions have been suggested before, 
   however broken by triplet superfields~\cite{PhysRevD.57.4174,PhysRevD.58.115007,PhysRevD.11.566}.  
Triplet Higgs superfields lead to a $U(1)_{em}$ violating vacuum~\cite{PhysRevD.48.4352, PhysRevLett.75.3989}. 
To keep a $U(1)_{em}$ invariant vacuum, at least one generation of right-handed scalar neutrino $\tilde{N}^c$ 
   must acquire a nonzero VEV. 
If we consider a supersymmetric LRM with the gauge group $SU(3)_c\times SU(2)_L\times SU(2)_R\times U(1)_{BL}$, 
   the the right-handed slepton doublet plays a role of the $SU(2)_R$ doublet Higgs field 
   and a VEV of right-handed scalar neutrino $\tilde{N}^c$ can break the LR symmetry down to the SM one \citep{FileviezPerez2009251}.   
It has been shown \cite{PhysRevLett.102.181802} that in the B-L extension of the minimal suspersymmetric Standard Model (MSSM), 
   the gauge group $SU(3)_c\times SU(2)_L\times U(1)_Y \times U(1)_{BL}$ is successfully broken down to 
   to the SM one by $\langle \tilde{N}^c \rangle$. 
In this context of the $U(1)_{BL}$ extension of the MSSM, radiative symmetry breaking can occur 
  when $\tilde{N}^c$'s mass squared becomes negative at low energies~\cite{PhysRevD.82.055002,1126-6708-2009-10-011}. 
Generally the seesaw mechanism comes about from a triplet scalar VEV inducing a Majorana mass term for the right-handed neutrino.  
However in this model, the seesaw is induced by the mixing between gaugino and neutrino~\cite{Barger:2010iv,Ghosh:2010hy}.

The main focus of this paper is to propose a class of supersymmetric LRMs, 
  where only a second Higgs bidoublet superfield is newly introduced, and 
  the LR symmetry is radiatively broken into the MSSM purely 
  by the VEV of the neutral component of right-handed slepton doublet. 
The LR symmetry breaking without any additional Higgs fields has been considered before~\cite{FileviezPerez2009251}, 
   where a negative mass squared for the right-handed slepton doublet is assumed. 
Here we calculate the renormalization group equations (RGEs) at the one-loop level and evolve them 
   from some intermediate scale down to the TeV scale. 
We find  that the mass squared of the right-handed slepton becomes negative 
   and hence the LR symmetry is radiatively broken. 
After the breaking, a charged lepton mixes with a charged gaugino, creating a sever bound on the gaugino mass 
   from the electroweak precision measurements. 
The neutral lepton component mixes with neutral gauginos and creates a heavy neutrino with a TeV scale mass.  
After EWSB the seesaw mechanism works to produce sub-eV scale neutrino masses. 
With the additional Higgs bidoublet, there are enough free parameters to reproduce realistic SM fermion mass matrices.

\section{Particle Content}

\begin{table}[t]
\begin{center}
\begin{tabular}{c|c|c|c|c}
& $SU(3)_c$ & $SU(2)_L$ & $SU(2)_R$ & $U(1)_{BL}$  \\
\hline
$Q=\left(\begin{array}{c}
u\\d
\end{array}\right) 
$ & ${\bf 3}$ & ${\bf 2}$ &${\bf 1}$&$1/3$ \\
$Q^c=\left(\begin{array}{c}
u^c\\d^c
\end{array}\right) $ &  $\bar{\bf 3} $& ${\bf 1}$ & ${\bf 2}$ & $-1/3$ \\
$L=\left(\begin{array}{c}
\nu\\e
\end{array}\right) $ & ${\bf 1}$ & ${\bf 2}$ & ${\bf 1}$ &$-1$\\
$L^c=\left(\begin{array}{c}
\nu^c\\e^c
\end{array}\right)$ &${\bf 1}$&${\bf 1}$&${\bf 2}$&$1$\\
$\Phi_i=\left(\begin{array}{c c}
\phi^+ & \phi^0_1\\ \phi^0_2 &  \phi^-
\end{array}\right) $& ${\bf 1}$ &${\bf 2}$&${\bf 2}$&$0$\\
\hline
\end{tabular}
\end{center}
\caption{
Particle content of our SUSY LR model. 
Two bidoubelt Higgs superfields $\Phi_i$ ($i=1,2$) are introduced. 
Here, we suppress the generation indices on the quark and lepton superfields
}\label{matter}
\end{table}

The particle content remains largely unchanged from the MSSM 
  as can bee seen in Table~\ref{matter}.
We extend the particle content in [18] by an extra Higgs bidoublet, which is necessary to obtain the realistic SM fermion mass matrices, 
  otherwise there is no flavor mixing in the model.   
The superpotential can be written down (flavor sums implied) as 
\begin{eqnarray}
\mathcal{W} &=& 
Y_q Q^T \tau_2 \Phi_1 \tau_2 Q^c +Y^\prime_q Q^T \tau_2\Phi_2 \tau_2 Q^c \nonumber \\ 
&+&Y_e L^T \tau_2\Phi_1\tau_2 L^c +Y^\prime_e L^T \tau_2\Phi_2 \tau_2L^c +\mu_{ii}\mathrm{Tr}\left(\Phi^T_i\tau_2\Phi_i\tau_2\right) \, ,
\label{SuperW}
\end{eqnarray} 
where we work the diagonal basis for the Higgs bidoublet without loss of generality.  
We can integrate a heavy Higgs bidoublet out at lower energies, and a lighter bidoublet 
   to be approximately identified as the MSSM Higgs.

The scalar potential with soft SUSY breaking masses is given by 
\begin{eqnarray} 
V_{soft} &=& 
  m^2_{\tilde{L}}|\tilde{L}|^2+m^2_{\tilde{L}^c}|\tilde{L}^c|^2+m^2_{\tilde{Q}}|\tilde{Q}|^2+m^2_{\tilde{Q}^c}|\tilde{Q}^c|^2 \\ \nonumber
  &+& m_{ij}^2\mathrm{Tr}\left(\Phi^\dagger_i \Phi_j\right)+B\mu_{ij}\mathrm{Tr}\left(\Phi^T_i\tau_2\Phi_j\tau_2\right)\, .
\label{ScalarV}
\end{eqnarray}
Here we have omitted $A$-terms, for simplicity, since their effects are not important in the following discussions. 
While the SUSY mass term for the two bidoublet Higgs superfields $\mu_{ij}$ is diagonal in Eq.~(\ref{SuperW}), 
  here we have introduced the off-diagonal $B\mu_{ij}$ term, 
  which will be tuned in order for the heavy Higgs bidoublet to develop a sizable VEV.

\section{RGE Analysis and Radiative LR symmetry breaking}
In our RGE analysis, we use a mixture of low energy data for the Standard Model gauge and Yukawa couplings 
  mixed with high energy inputs inspired by the MSSM. 
For Yukawa couplings we only consider the 3rd generation. 
Using the RGEs of the SM \cite{PhysRevD.46.3945} 
  at the one-loop level we run them from $\mu=M_Z$ to $\mu=$ 1 TeV. 
Taking the outputs of the previous SM RGE runnings at $\mu=1$ TeV as inputs for the RGEs of the MSSM \cite{PhysRevD.49.4882} 
  at the one-loop level, we solve the MSSM RGEs until LR symmetry breaking scale $v_R$. 
In this paper, we fix $v_R = 20$ TeV  as a reference value. 
At the one-loop level the soft mass terms do not affect the runnings of the gauge and Yukawa couplings. 
At the LR symmetry breaking we have the relations between the hypercharge gauge coupling ($g_Y$) 
  and the LR gauge couplings ($g_R$ and $g_{BL}$) as 
\begin{eqnarray}\label{rel}
  g_Y=g_R\sin{\theta_R}\, ,\quad\tan{\theta_R}=2 \;  \frac{g_{BL}}{g_R}\, .
\end{eqnarray}
In this analysis we choose, for simplicity, $\theta_R=65^\circ$, $g_{BL}=0.438$, and $g_{R}=0.408$, 
  which are evaluated at $v_R=20$ TeV based on Eq.~(\ref{rel}) from the known MSSM gauge couplings. 
The values of the tau and top Yukawa couplings from the MSSM RGEs at $\mu= 20$ TeV are evaluated 
   as $Y_\tau \simeq 0.01$ and $Y_t \simeq 0.8$.  
As a matter of simplicity we choose $Y_{q}=0.7 \; Y_t$ and $Y_{q'}=0.3 \,Y_t$ and $Y_{l'}=Y_l=Y_\tau/2$ as inputs at $\mu= 20$ TeV. 
We run the RGEs for the Yukawa couplings and gauge couplings (see Eqs.~(\ref{gauge})-(\ref{by}) in Appendix A) 
   from 20 TeV up to a SUSY breaking mediation scale which we choose to be an intermediate scale $\mu=10^{12}$ GeV, 
   for simplicity.   
At the scale of $10^{12}$ GeV, we take all gaugino masses to be $2.5$ TeV except for 
    the $SU(2)_R$ gaugino which is 100 TeV to keep the guagino-lepton mixing within the current experimental bound. 
This bound will be discussed below. 
The RGE invariant relation in Eq.~(\ref{gaugem}) is used for the gaugino masses. 
We calculate the RGE evolutions in Eqs.~(\ref{fir})-(\ref{las}) for the soft masses at the one-loop level 
   and run them down from $\mu=10^{12}$ GeV to $\mu=20$ TeV.  
We use the evaluated Yukawa and gauge couplings at $\mu=10^{12}$ GeV as inputs into the soft mass RGEs. 
To realize the LR symmetry breaking the non-universal soft mass inputs are crucial. 
See Table~\ref{table2} for our inputs at $\mu=10^{12}$ GeV and outputs at $\mu=20$ TeV. 

Our choices for the masses are a result of straightforward numerical calculation of RGEs.  
At $\mu=10^{12}$ GeV, $g_{BL}$ is the largest coupling so Yukawas can be ignored except for the RGEs 
 for the bidoublet Higgs mass squares. 
Because of this size, the sign in front of the D-term trace given in Eq.~(\ref{dtrace}),  
  which is involved in the RGEs of Eqs.~(\ref{fir})-(\ref{las}), 
  will dominate and could drive the soft mass square  of  $\tilde{L}^c$ negative at low energies.

{\renewcommand{\arraystretch}{1.2}
\begin{table}[t]
\begin{center}
\begin{tabular}{|c|c|c|}
\hline
 & $\mu=20$ TeV & $\mu=10^{12}$  TeV \\  \hline
$ M_{\tilde{L}_1^c}^2 $& $ 2.0 \times 10^9\, \text{GeV}^2 $ & $2.5\times 10^9\, \text{GeV}^2$ \\\hline
$ M_{\tilde{L}_2^c}^2 $& $2.0 \times 10^9\, \text{GeV}^2$ & $2.5\times 10^9 \,\text{GeV}^2$ \\\hline
$ M_{\tilde{L}_3^c}^2 $ & $ -4.7\times 10^7\, \text{GeV}^2 $ & $ 2.1\times 10^4  \,\text{GeV}^2$ \\\hline
$ M_{\tilde{Q}_3^c}^2 $& $3.1\times 10^9 \,\text{GeV}^2 $ & $2.5\times 10^9 \,\text{GeV}^2$ \\\hline
$ M_{\tilde{Q}_3}^2 $& $1.3\times 10^{10} \,\text{GeV}^2 $ & $1.4\times 10^{10} \,\text{GeV}^2 $ \\\hline
$ M_{\tilde{L}_3}^2 $&  $4.2\times 10^9\, \text{GeV}^2 $ & $2.5\times 10^9 \,\text{GeV}^2 $ \\\hline
$ M^2_{\Phi _1} $& $1.0\times 10^6 \,\text{GeV}^2 $ & $2.1\times 10^8 \,\text{GeV}^2 $ \\\hline
$ M^2_{\Phi _2} $& $3.4\times 10^9\, \text{GeV}^2 $ & $2.5\times 10^9 \,\text{GeV}^2 $ \\\hline
$ M_{\tilde g} $  & $5000\, \text{GeV} $ & $2500\, \text{GeV}$ \\ \hline  
$ M_L $& $2300\, \text{GeV} $ & $2500\, \text{GeV}$ \\\hline
$ M_R$ & $10^5\, \text{GeV}$ & $10^5 \,\text{GeV}$ \\\hline
$ M_{\text{BL}} $ & $ 800\, \text{GeV} $ & $2500 \text{GeV} $ \\\hline
\end{tabular}
\end{center}
\caption{
List of soft masses at $\mu=10^{12}$ GeV (inputs) and at $\mu=20$ TeV (outputs). 
$M_{\tilde{g}}$, $M_L$, $M_R$ and $M_{BL}$ are gaugino masses 
  corresponding to $SU(3)_c$, $SU(2)_L$, $SU(2)_R$ and $U(1)_{BL}$, respectively. 
}
\label{table2}
\end{table}

\begin{figure}[t]
\centering
\includegraphics[width=10cm]{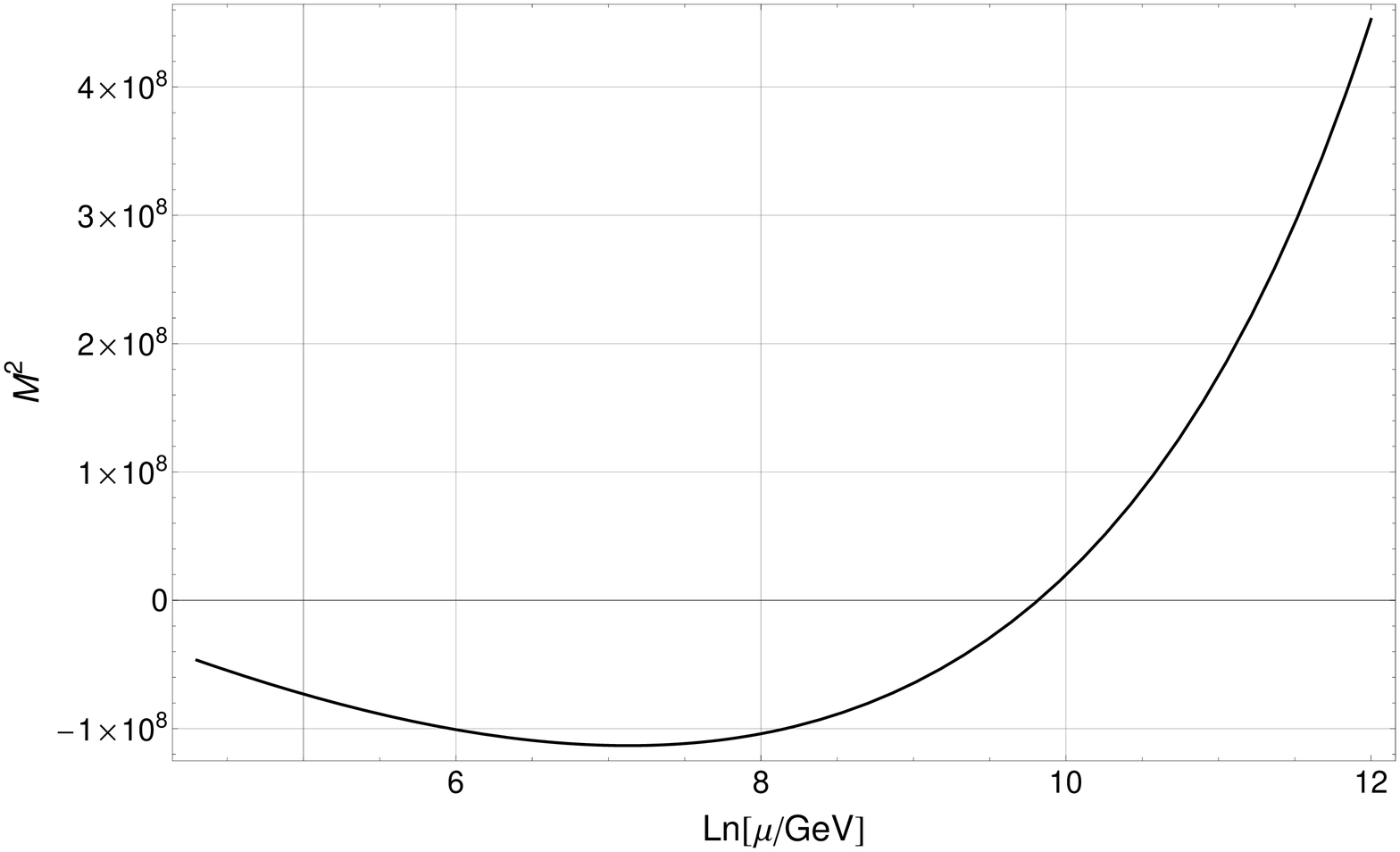}
\caption{
The RGE evolution of the soft mass squared for $\tilde{L}^c_3$, 
  which becomes negative at low energies. 
}
\label{fig:sleptons}
\end{figure}

The running mass squared for $\tilde{L}^c_3$ is shown in Fig \ref{fig:sleptons}. 
We see that it becomes negative at low energies. 
Here we consider the case that the 3rd generation right-handed slepton doublet 
  acquires the negative mass squared. 
The potential for $\tilde{L}^c_3$ is described as 
\begin{equation}
V= m^2_{\tilde{L}^c_3} |{\tilde{L}^c_3} |^2 +\frac{1}{8}  \left( g^2_R+4g_{BL}^2 \right) |{\tilde{L}^c_3} |^4   \; , 
\end{equation}
  and the right-handed scalar neutrino $\tilde{N}^c_3$ develops its VEV at the potential minimum 
  as $\langle \tilde{N}^c_3 \rangle = v_R/\sqrt{2}$, where 
\begin{equation}
v_R=\sqrt{\frac{-8m^2_{\tilde{L}^c_3}}{g^2_R+4g^2_{BL}}}\, .
\end{equation}
The numerical value in this model for the VEV is 20 TeV and $m^2_{\tilde{L}^c_1}$ is evaluated at 20 TeV. 
Since the $SU(2)_R\times U(1)_{BL}$ symmetry is broken by the $SU(2)_R$ doublet VEV, 
   the gauge boson mass relations are very similar to those in the SM. 
One gauge boson remains massless which is identified as the $U(1)_Y$ gauge boson 
  while the three massive ones and a charge relation are 
\begin{eqnarray}
M_{W_R}&=&\frac{1}{2}g_Rv_R\, ,\\
M_{Z_R}&=&\frac{1}{2}\sqrt{g_R^2+4g^2_{BL}}v_R\, ,\\
Q_Y &=&\frac{Q_{BL}}{2}- T_R^3 \,.
\end{eqnarray}
The gauge boson masses based on our runnings of the couplings and above VEV come out to be 4.1 TeV and 9.6 TeV, respectively, 
   which satisfies the LHC bound of $M_{W_R} \gtrsim 3$ TeV~\cite{wrexp}.

\section{Mass bound on $SU(2)_R$ gaugino} 
In the above, we stated that there is a bound on the $SU(2)_R$ gaugino mass.  
This bound is unique to this model where the LR symmetry is broken by the VEV of right-handed neutrino. 
After the breaking of the LR symmetry, 
  the right-handed tau is mixed with the $SU(2)_R$ gaugino. 
The relevant terms are 
\begin{equation}
\mathcal{L}\supset M_R \tilde{\lambda}^+\tilde{\lambda}^-+\frac{1}{\sqrt{2}}g_R v_R \tilde{\lambda}^- E^c= M_R \tilde{\lambda}^+\tilde{\lambda}^-+\sqrt{2}M_{W_R} \tilde{\lambda}^- E^c\,.
\end{equation}
We diagonalize the mass matrix as 
\begin{align}
\xi_1^+= \cos{\phi}\tilde{\lambda^+_R}+\sin{\phi}E^c\qquad \text{and}\qquad  \xi_2^+= \cos{\phi}E^c-\sin{\phi}\tilde{\lambda^+_R}\, 
\end{align}
with a mixing angle
\begin{equation}
\tan{\phi}=\frac{\sqrt{2}M_{W_R}}{M_R}\,.
\label{taumix}
\end{equation}
The neutral current for the charged leptons in the SM is now modified as
\begin{equation}
\label{nj}
J^\mu_Z=\frac{2m_Z}{v}\left[
 \left(-\frac{1}{2}+\sin^2 \theta_W \right) \overline{\tau_L}\gamma^\mu \tau_L+ 
 \sin^2 \theta_W \cos^2{\phi} \; \overline{\tau_R} \gamma^\mu \tau_R 
 \right]\,,
\end{equation}
where $v=246$ GeV,  $\theta_W$ is the weak mixing angle, and $m_Z=91.2$ GeV. 
Using the precision data at the LEP experiment for $Z\rightarrow \tau^+\tau^-$ decay width uncertainties, 
  the modification of the weak neutral current must not change the width by more than $|\delta \Gamma| = 0.22$ MeV \cite{Zprecision}.  
Using Eq.~(\ref{nj}), we calculate the change of the decay width as
\begin{equation}
\delta \Gamma=\frac{m_Z^3 \sin^4 \theta_W}{6\pi v^2}(\cos^4 \phi-1)\approx -\frac{m_Z^3 \sin^4 \theta_W}{6\pi v^2}\left(\frac{ 4M^2_{W_R}}{M_R^2}\right), 
\end{equation}   
where we have used Eq.~(\ref{taumix}) and $|\phi| \ll 1$. 
Now we interpret the LEP bound as $M_R \gtrsim 25M_{W_R}$. 
At the scale of $v_R$=20 TeV we calculate $M_{W_R}=$4.1 TeV, so the mass $M_R= 100$ TeV  
   shown in Table~\ref{table2} is consistent with the LEP bound.

\section{SM fermion mass matrices}
We first examine the neutral  fermion sector to analyze the mixing between the gauginos and leptons 
  from the SUSY  gauge interaction after $\tilde{L}^c$ develops a nonzero VEV.  
The hypercharge $Q_Y=0$ sector of the Lagrangian after LR symmetry breaking is
\begin{equation}
\mathcal{L}\supset g_{BL}v_R\nu^c\lambda_{BL}+ \frac{1}{2}g_{R}v_R\nu^c\lambda^3_{R}+\frac{1}{2}M_R\lambda^3_{R}\lambda^3_{R}+\frac{1}{2}M_{BL}\lambda_{BL}\lambda_{BL}, 
\end{equation}
where $\lambda^3_{R}$ is the gaugino corresponding to the $SU(2)_R$ generator $T_R^3$. 
The mass matrix after the LR symmetry breaking is found to be 
\begin{equation}M_{\tilde{\lambda}^3_{R}, \tilde{\lambda}_{BL},\nu^c}
=
\left(
\begin{array}{c c c}
M_R & 0 & \frac{1}{2}g_R v_R\\
0 & M_{BL} & g_{BL}v_R\\
\frac{1}{2} g_R v_R & g_{BL}v_R & 0
\end{array}
\right).
\end{equation}  
Because of the LEP bound $M_R \gg M_{W_R}$, $\lambda_R^3$ is decoupled, 
  while the right-handed neutrino ($\nu^c$) acquires its Majorana mass of ${\cal O}$(1 TeV) 
  through the mixing with the B-L gaugino with $M_{BL}$, $g_R v_R$, $g_{BL} v_R={\cal O}$(1 TeV).  
With this right-handed neutrino mass of ${\cal O}$(1 TeV), the seesaw mechanism works in our model.

After EWSB, the SM fermion mass matrices can be expressed as 
\begin{align}
M_t&=\frac{1}{\sqrt{2}}Y_Q v_u+\frac{1}{\sqrt{2}}Y^\prime_Q v^\prime_u=M_Q+M_Q^\prime\, ,\\
\label{mb}M_b&=\frac{1}{\sqrt{2}}Y_Q v_d+\frac{1}{\sqrt{2}}Y^\prime_Q v^\prime_d=cM_Q+c'M_Q^\prime\, ,\\
\label{mn}M^D_\nu&=\frac{1}{\sqrt{2}}Y_L v_u+\frac{1}{\sqrt{2}}Y^\prime_L v^\prime_u=M_L+M_L^\prime\, ,\\
M_\tau&=\frac{1}{\sqrt{2}}Y_L v_d+\frac{1}{\sqrt{2}}Y^\prime_L v^\prime_d=cM_L+c'M_L^\prime\, ,
\end{align}
where $c=v_d/v_u$ and $c'=v'_d/v'_u$, and we have considered the 3rd generation to simplify our discussion. 
Since there are two Higgs bidoublets creating four nonzero VEVs, they can all be paramterized on a 4-sphere, 
  allowing for 3 free parameters under the constraint $v_u^2+v_d^2+v^{\prime 2}_u+v^{\prime 2}_d=(246)^2$ GeV$^2$.  
We tune $Y_L^\prime$ so that there is a cancellation in Eq.~\eqref{mn} to produce the neutrino Dirac mass, 
  $M^D_\nu=\mathcal{O}(10^{-3}$ GeV), 
  while allowing for the tau lepton Dirac mass $M_\tau=\mathcal{O}(1\,\text{GeV})$. 
In the quark sector we tune the quark Yukawa coupling, $Y^\prime_Q$, so that there is a cancellation 
   in Eq.~\eqref{mb} to produce $M_b=\mathcal{O}(1$ GeV) while the top quark mass equation produces $M_t=\mathcal{O}(100$ GeV). 
Our discussion here is easily extended to the three generation case, and we can reproduce realistic SM fermion mass matrices.

The Dirac mass term for the neutrinos will further mix with the Higgsinos and neutral gauginos from the EW sector 
  as well to produce a neutralino mass matrix 
\begin{equation}\label{neu}
\left(
\begin{array}{cccccccc}
0 & \mu_{11} & 0 & 0 & Y_L\frac{v_R}{\sqrt{2}} & 0 & 0 & 0\\
\mu_{11} & 0 & 0 & 0 & 0 & 0 & 0 & 0\\
0 & 0 & 0 & \mu_{22} & Y^\prime_L\frac{v_R}{\sqrt{2}}& 0 & 0 & 0\\
0 & 0 & \mu_{22} & 0 & 0 & 0 & 0 & 0\\
Y_L\frac{v_R}{\sqrt{2}}& 0 &Y^\prime_L\frac{v_R}{\sqrt{2}} & 0 & 0 & M^D_\nu & 0 & 0\\
0 & 0 & 0 & 0 &M^D_\nu & 0 & M_{W_R}\tan \theta_R & M_{W_R}\\
0 & 0 & 0 & 0 & 0 & M_{W_R}\tan \theta_R & M_{BL} & 0\\
0 & 0 & 0 & 0 & 0 & M_{W_R} & 0 & M_R\\
\end{array}
\right)\, .
\end{equation}
For simplicity we took the one generation case. 
This can be easily extended to the 3 generation case by promoting the Yukawa couplings to $3 \times 3$ matrices. 
Since $M_R \gg M_{W_R}$, the $SU(2)_R$ gaugino is decoupled. 
To understand the seesaw mechanism in our model, we focus on the block-diagonal $3\times3$ matrix 
  composed of the elements $M_\nu^D$, $M_{W_R} \tan \theta_R$ and $M_{BL}$.  
Since $M_{W_R} \tan \theta_R$, $M_{BL}={\cal O}$(1 TeV)$\gg M^D_\nu ={\cal O}$(1 MeV),   
  we find a mass eigenvalue for the light neutrino as 
\begin{equation}
  m_\nu \simeq \frac{\left(M^D_\nu\right)^2}{M_{BL}} ={\cal O}(0.1 \; {\rm eV})
\end{equation}
 through the seesaw mechanism.\footnote{
It is interesting to notice that if $M_{BL}\gg M_{W_R}$ the block-diagonal matrix 
 has a ``double seesaw'' structure, leading to mass eigenvalues approximately given by 
 $(M_\nu^D)^2/{\tilde M}$, ${\tilde M}\simeq (M_{W_R} \tan \theta_R)^2/M_{BL}$ and $M_{BL}$. 
}

\section{Conclusions}
We have considered a SUSY Left-Right symmetric model  
   based on the gauge group $SU(3)_c\times SU(2)_L\times SU(2)_R\times U(1)_{BL}$,  
   where in addition to the quark and lepton superfields only two Higgs bidoublets are introduced. 
With suitable soft mass inputs at a  SUSY breaking mediation scale, where scalar squared masses are all positive,  
   we have found that a right-handed slepton doublet mass squared becomes negative in its RG evolution, 
   and as a result, the LR symmetry is radiatively broken to the SM gauge group by a right-handed neutrino VEV.  
The right-handed neutrino VEV also generates a mass mixing between the $SU(2)_R$ gaugino and SM right-handed lepton.  
This is a unique feature of our model, and the mass mixing is severely constrained by the LEP electroweak precision data. 
We have found the mass ratio of $M_R \gtrsim 25 M_{W_R}$ from the LEP bound.  
Realistic SM fermion mass matrices can be reproduced by the introduction of the two Higgs bidoublets 
  and suitable tunings of Yukawa matrices. 
The right-handed neutrinos acquire Majorana masses of ${\cal O}$(1 TeV) through its mixing with the B-L gaugino, 
  and the seesaw mechanism works to generate a light neutrino mass of sub-eV scale.
  
In our model, $R$-parity is also broken by the right-handed sneutrino VEV,
  so that the lightest superpartner (LSP) neutralino, which is the conventional dark matter candidate in SUSY models,
  becomes unstable and no longer remains a viable dark matter candidate.
As discussed in \cite{RefA,RefB}, even in the presence of $R$-parity violation, an unstable gravitino if it is the LSP
  has a lifetime longer than the age of the universe and can still be the dark matter candidate.
Hence, as a simple way to incorporate a dark matter candidate in our model, we can consider the LSP gravitino scenario.
However, with the given mass hierarchy $M_R=100$ TeV $\gg M_{BL}=800$ GeV, it is difficult to naturally provide
  the LSP gravitino in 4-dimensional supergravity mediated SUSY breaking.
For a simple realization, we may consider a gravity mediated SUSY breaking in a warped 5-dimensional supergravity~\cite{IOY},
  where gravitino is always the LSP with a SUSY breaking mediation scale being ``warped down" from the Planck mass.
 This gravity mediation at low energies fits the choice of the SUSY breaking mediation scale to be $\mu=10^{12}$ GeV
  in our RGE analysis.

\appendix
\section{Renormalization Group Equations}
The RGEs for the gauge couplings are
\begin{eqnarray}\label{gauge}
16\pi^2 \frac{d\,g_{i}}{d(\ln\mu)}&=b_ig^3_{i}\, ,
\end{eqnarray}
The gaugino masses can be simply defined using the RGE invariant quantity 
\begin{eqnarray}\label{gaugem}
\frac{d}{d(\ln\mu)}\left(\frac{M_i}{g_i^2}\right)=0\,.
\end{eqnarray}
RGEs for the Yukawa couplings at the  one-loop level are described as 
\begin{align}
16\pi^2 \frac{d \mathbf{Y}_i}{d(\ln\mu)}=\mathbf{Y}_i\beta_{i}\, ,
\end{align}
where the beta functions for each Yukawa are defined as 
\begin{eqnarray}
\label{bq}
\beta_q &=&4\mathbf{Y}^\dagger_q \mathbf{Y}_q+\rm{Tr}[3\mathbf{Y}^\dagger_q \mathbf{Y}_q+\mathbf{Y}^\dagger_l \mathbf{Y}_l+(3\mathbf{Y}_q^\dagger \mathbf{Y}'_q+\mathbf{Y}_l^\dagger \mathbf{Y}'_l+h.c.)]\nonumber\\
&-&\left(\frac{4}{9}g^2_{BL}+3g^2_R+3g^2_L+\frac{16}{3}g^2_3\right),\\
\beta_{q'} &=&4\mathbf{Y'}^\dagger_q \mathbf{Y'}_q+\rm{Tr}[3\mathbf{Y'}^\dagger_q \mathbf{Y'}_q+\mathbf{Y'}^\dagger_l \mathbf{Y'}_l+(3\mathbf{Y}_q^\dagger \mathbf{Y}'_q+\mathbf{Y}_l^\dagger \mathbf{Y}'_l+h.c.)]\nonumber\\
&-&\left(\frac{4}{9}g^2_{BL}+3g^2_R+3g^2_L+\frac{16}{3}g^2_3\right),\\
\beta_l &=&4\mathbf{Y}^\dagger_l \mathbf{Y}_l+\rm{Tr}[3\mathbf{Y}^\dagger_q \mathbf{Y}_q+\mathbf{Y}^\dagger_l \mathbf{Y}_l+(3\mathbf{Y}_q^\dagger \mathbf{Y}'_q+\mathbf{Y}_l^\dagger \mathbf{Y}'_l+h.c.)]\nonumber\\
&-&\left(4g^2_{BL}+3g^2_R+3g^2_L\right), 
\\
\label{by}\beta_{l'} &=&4\mathbf{Y'}^\dagger_l \mathbf{Y'}_l+\rm{Tr}[3\mathbf{Y'}^\dagger_q \mathbf{Y'}_q+\mathbf{Y'}^\dagger_l \mathbf{Y'}_l+(3\mathbf{Y}_q^\dagger \mathbf{Y}'_q+\mathbf{Y}_l^\dagger \mathbf{Y}'_l+h.c.)]\nonumber\\
&-&\left(4g^2_{BL}+3g^2_R+3g^2_L\right)\, .
\end{eqnarray}
The soft mass RGEs are
\begin{eqnarray}
\label{fir}
8\pi^2 \frac{d m^2_{\tilde{Q}_i}}{d(\ln\mu)}&=&\sum_{j,k} |Y_{Q}^{ijk}|^2\left(m^2_{\tilde{Q}_i}+m^2_{\tilde{Q}^c_j}+m^2_{\Phi_k}\right)\nonumber\\
&+&\frac{1}{3}g_{BL}^2\mathrm{Tr}[Q_{BL}m^2]-\frac{4}{9}g_{BL}^2M_{BL}^2-3g_{L}^2 M^2_{L}-\frac{16}{3}g^2_3M^2_3 \; ,\\
8\pi^2 \frac{d m^2_{\tilde{Q}^c_i}}{d(\ln\mu)}&=&\sum_{j,k} |Y_{Q}^{ijk}|^2\left(m^2_{\tilde{Q}_i}+m^2_{\tilde{Q}^c_j}+m^2_{\Phi_k}\right) \nonumber\\
&-&\frac{1}{3}g_{BL}^2\mathrm{Tr}[Q_{BL}m^2]-\frac{4}{9}g_{BL}^2M_{BL}^2-3g_{R}^2 M^2_{R}-\frac{16}{3}g^2_3M^2_3 \; ,\\
8\pi^2\frac{d m^2_{\tilde{L_i}}}{d(\ln\mu)} &=&\sum_{j,k} |Y_{L}^{ijk}|^2\left(m^2_{\tilde{L}_i}+m^2_{\tilde{L}^c_j}+m^2_{\Phi_k}\right) \nonumber\\
&-& g_{BL}^2\mathrm{Tr}[Q_{BL}m^2]-4g_{BL}^2M_{BL}^2-3g_{L}^2 M^2_{L} \; ,\\
\label{msl}8\pi^2\frac{d m^2_{\tilde{L_i}^c}}{d(\ln\mu)} &=&\sum_{j,k} |Y_{L}^{ijk}|^2\left(m^2_{\tilde{L}_i}+m^2_{\tilde{L}^c_j}+m^2_{\Phi_k}\right) \nonumber\\
&+& g_{BL}^2\mathrm{Tr}[Q_{BL}m^2]-4g_{BL}^2M_{BL}^2-3g_{R}^2 M^2_{R} \; ,\\
8\pi^2\frac{d m^2_{\Phi_1}}{d(\ln\mu)} &=&3\sum_{i,j} |Y_{Q}^{ij}|^2\left(m^2_{\tilde{Q}_i}+m^2_{\tilde{Q}^c_j}+m^2_{\Phi_1}\right)+\sum_{i,j} |Y_{L}^{ij}|^2\left(m^2_{\tilde{L}_i}+m^2_{\tilde{L}^c_j}+m^2_{\Phi_1}\right)\nonumber\\
&-& 3g_{L}^2 M^2_{L}-3g_{R}^2 M^2_{R} \; , \\
\label{las}8\pi^2\frac{d m^2_{\Phi_2}}{d(\ln\mu)} &=&3\sum_{i,j} |Y_{Q}^{\prime \,ij}|^2\left(m^2_{\tilde{Q}_i}+m^2_{\tilde{Q}^c_j}+m^2_{\Phi_2}\right)+\sum_{i,j} |Y_{L}^{\prime \,ij}|^2\left(m^2_{\tilde{L}_i}+m^2_{\tilde{L}^c_j}+m^2_{\Phi_2}\right)\nonumber\\
&-& 3g_{L}^2 M^2_{L}-3g_{R}^2 M^2_{R}\, .
\end{eqnarray}
For equations \eqref{fir}-\eqref{las}, the trace terms are defined as 
\begin{eqnarray}\label{dtrace} \mathrm{Tr}\left[Q_{BL}m^2\right]=
  2\sum_i \left(m^2_{\tilde{Q}_i}-m^2_{\tilde{Q}_i^c}-m^2_{\tilde{L}_i}+m^2_{\tilde{L}_i^c}\right)\,.
\end{eqnarray}

\bibliographystyle{aipauth4-1}

\bibliography{science}

\begin{thebibliography}{24}%
\makeatletter
\providecommand \@ifxundefined [1]{%
 \@ifx{#1\undefined}
}%
\providecommand \@ifnum [1]{%
 \ifnum #1\expandafter \@firstoftwo
 \else \expandafter \@secondoftwo
 \fi
}%
\providecommand \@ifx [1]{%
 \ifx #1\expandafter \@firstoftwo
 \else \expandafter \@secondoftwo
 \fi
}%
\providecommand \natexlab [1]{#1}%
\providecommand \enquote  [1]{``#1''}%
\providecommand \bibnamefont  [1]{#1}%
\providecommand \bibfnamefont [1]{#1}%
\providecommand \citenamefont [1]{#1}%
\providecommand \href@noop [0]{\@secondoftwo}%
\providecommand \href [0]{\begingroup \@sanitize@url \@href}%
\providecommand \@href[1]{\@@startlink{#1}\@@href}%
\providecommand \@@href[1]{\endgroup#1\@@endlink}%
\providecommand \@sanitize@url [0]{\catcode `\\12\catcode `\$12\catcode
  `\&12\catcode `\#12\catcode `\^12\catcode `\_12\catcode `\%12\relax}%
\providecommand \@@startlink[1]{}%
\providecommand \@@endlink[0]{}%
\providecommand \url  [0]{\begingroup\@sanitize@url \@url }%
\providecommand \@url [1]{\endgroup\@href {#1}{\urlprefix }}%
\providecommand \urlprefix  [0]{URL }%
\providecommand \Eprint [0]{\href }%
\providecommand \doibase [0]{http://dx.doi.org/}%
\providecommand \selectlanguage [0]{\@gobble}%
\providecommand \bibinfo  [0]{\@secondoftwo}%
\providecommand \bibfield  [0]{\@secondoftwo}%
\providecommand \translation [1]{[#1]}%
\providecommand \BibitemOpen [0]{}%
\providecommand \bibitemStop [0]{}%
\providecommand \bibitemNoStop [0]{.\EOS\space}%
\providecommand \EOS [0]{\spacefactor3000\relax}%
\providecommand \BibitemShut  [1]{\csname bibitem#1\endcsname}%
\let\auto@bib@innerbib\@empty
\bibitem [{\citenamefont {Pati}\ and\ \citenamefont
  {Salam}(1974)}]{PhysRevD.10.275}%
  \BibitemOpen
  \bibfield  {author} {\bibinfo {author} {\bibfnamefont {J.~C.}\ \bibnamefont
  {Pati}}\ and\ \bibinfo {author} {\bibfnamefont {A.}~\bibnamefont {Salam}},\
  }\href {\doibase 10.1103/PhysRevD.10.275} {\bibfield  {journal} {\bibinfo
  {journal} {Phys. Rev. D}\ }\textbf {\bibinfo {volume} {10}},\ \bibinfo
  {pages} {275} (\bibinfo {year} {1974})}\BibitemShut {NoStop}%
\bibitem [{\citenamefont {Mohapatra}\ and\ \citenamefont
  {Pati}(1975{\natexlab{a}})}]{PhysRevD.11.566}%
  \BibitemOpen
  \bibfield  {author} {\bibinfo {author} {\bibfnamefont {R.~N.}\ \bibnamefont
  {Mohapatra}}\ and\ \bibinfo {author} {\bibfnamefont {J.~C.}\ \bibnamefont
  {Pati}},\ }\href {\doibase 10.1103/PhysRevD.11.566} {\bibfield  {journal}
  {\bibinfo  {journal} {Phys. Rev. D}\ }\textbf {\bibinfo {volume} {11}},\
  \bibinfo {pages} {566} (\bibinfo {year} {1975}{\natexlab{a}})}\BibitemShut
  {NoStop}%
\bibitem [{\citenamefont {Khachatryan}\ \emph {et~al.}(2014)\citenamefont
  {Khachatryan} \emph {et~al.}}]{wrexp}%
  \BibitemOpen
  \bibfield  {author} {\bibinfo {author} {\bibfnamefont {V.}~\bibnamefont
  {Khachatryan}} \emph {et~al.},\ }\href {\doibase
  10.1140/epjc/s10052-014-3149-z} {\bibfield  {journal} {\bibinfo  {journal}
  {The European Physical Journal C}\ }\textbf {\bibinfo {volume} {74}},\
  \bibinfo {eid} {3149} (\bibinfo {year} {2014}),\
  10.1140/epjc/s10052-014-3149-z}\BibitemShut {NoStop}%
\bibitem [{\citenamefont {Maiezza}\ \emph {et~al.}(2010)\citenamefont
  {Maiezza}, \citenamefont {Nemev\ifmmode~\check{s}\else \v{s}\fi{}ek},
  \citenamefont {Nesti},\ and\ \citenamefont {Senjanovi\ifmmode~\acute{c}\else
  \'{c}\fi{}}}]{PhysRevD.82.055022}%
  \BibitemOpen
  \bibfield  {author} {\bibinfo {author} {\bibfnamefont {A.}~\bibnamefont
  {Maiezza}}, \bibinfo {author} {\bibfnamefont {M.}~\bibnamefont
  {Nemev\ifmmode~\check{s}\else \v{s}\fi{}ek}}, \bibinfo {author}
  {\bibfnamefont {F.}~\bibnamefont {Nesti}}, \ and\ \bibinfo {author}
  {\bibfnamefont {G.}~\bibnamefont {Senjanovi\ifmmode~\acute{c}\else
  \'{c}\fi{}}},\ }\href {\doibase 10.1103/PhysRevD.82.055022} {\bibfield
  {journal} {\bibinfo  {journal} {Phys. Rev. D}\ }\textbf {\bibinfo {volume}
  {82}},\ \bibinfo {pages} {055022} (\bibinfo {year} {2010})}\BibitemShut
  {NoStop}%
\bibitem [{\citenamefont {Mohapatra}\ and\ \citenamefont
  {Pati}(1975{\natexlab{b}})}]{PhysRevD.11.2558}%
  \BibitemOpen
  \bibfield  {author} {\bibinfo {author} {\bibfnamefont {R.~N.}\ \bibnamefont
  {Mohapatra}}\ and\ \bibinfo {author} {\bibfnamefont {J.~C.}\ \bibnamefont
  {Pati}},\ }\href {\doibase 10.1103/PhysRevD.11.2558} {\bibfield  {journal}
  {\bibinfo  {journal} {Phys. Rev. D}\ }\textbf {\bibinfo {volume} {11}},\
  \bibinfo {pages} {2558} (\bibinfo {year} {1975}{\natexlab{b}})}\BibitemShut
  {NoStop}%
\bibitem [{\citenamefont {Senjanovic}\ and\ \citenamefont
  {Mohapatra}(1975)}]{PhysRevD.12.1502}%
  \BibitemOpen
  \bibfield  {author} {\bibinfo {author} {\bibfnamefont {G.}~\bibnamefont
  {Senjanovic}}\ and\ \bibinfo {author} {\bibfnamefont {R.~N.}\ \bibnamefont
  {Mohapatra}},\ }\href {\doibase 10.1103/PhysRevD.12.1502} {\bibfield
  {journal} {\bibinfo  {journal} {Phys. Rev. D}\ }\textbf {\bibinfo {volume}
  {12}},\ \bibinfo {pages} {1502} (\bibinfo {year} {1975})}\BibitemShut
  {NoStop}%
\bibitem [{\citenamefont {Minkowski}(1977)}]{Minkowski1977421}%
  \BibitemOpen
  \bibfield  {author} {\bibinfo {author} {\bibfnamefont {P.}~\bibnamefont
  {Minkowski}},\ }\href {\doibase
  http://dx.doi.org/10.1016/0370-2693(77)90435-X} {\bibfield  {journal}
  {\bibinfo  {journal} {Physics Letters B}\ }\textbf {\bibinfo {volume} {67}},\
  \bibinfo {pages} {421 } (\bibinfo {year} {1977})}\BibitemShut {NoStop}%
\bibitem [{\citenamefont {Deshpande}\ \emph {et~al.}(1991)\citenamefont
  {Deshpande}, \citenamefont {Gunion}, \citenamefont {Kayser},\ and\
  \citenamefont {Olness}}]{PhysRevD.44.837}%
  \BibitemOpen
  \bibfield  {author} {\bibinfo {author} {\bibfnamefont {N.~G.}\ \bibnamefont
  {Deshpande}}, \bibinfo {author} {\bibfnamefont {J.~F.}\ \bibnamefont
  {Gunion}}, \bibinfo {author} {\bibfnamefont {B.}~\bibnamefont {Kayser}}, \
  and\ \bibinfo {author} {\bibfnamefont {F.}~\bibnamefont {Olness}},\ }\href
  {\doibase 10.1103/PhysRevD.44.837} {\bibfield  {journal} {\bibinfo  {journal}
  {Phys. Rev. D}\ }\textbf {\bibinfo {volume} {44}},\ \bibinfo {pages} {837}
  (\bibinfo {year} {1991})}\BibitemShut {NoStop}%
\bibitem [{\citenamefont {Aulakh}\ \emph
  {et~al.}(1998{\natexlab{a}})\citenamefont {Aulakh}, \citenamefont {Melfo},\
  and\ \citenamefont {Senjanovi\ifmmode~\acute{c}\else
  \'{c}\fi{}}}]{PhysRevD.57.4174}%
  \BibitemOpen
  \bibfield  {author} {\bibinfo {author} {\bibfnamefont {C.~S.}\ \bibnamefont
  {Aulakh}}, \bibinfo {author} {\bibfnamefont {A.}~\bibnamefont {Melfo}}, \
  and\ \bibinfo {author} {\bibfnamefont {G.}~\bibnamefont
  {Senjanovi\ifmmode~\acute{c}\else \'{c}\fi{}}},\ }\href {\doibase
  10.1103/PhysRevD.57.4174} {\bibfield  {journal} {\bibinfo  {journal} {Phys.
  Rev. D}\ }\textbf {\bibinfo {volume} {57}},\ \bibinfo {pages} {4174}
  (\bibinfo {year} {1998}{\natexlab{a}})}\BibitemShut {NoStop}%
\bibitem [{\citenamefont {Aulakh}\ \emph
  {et~al.}(1998{\natexlab{b}})\citenamefont {Aulakh}, \citenamefont {Melfo},
  \citenamefont {Ra\ifmmode~\check{s}\else \v{s}\fi{}in},\ and\ \citenamefont
  {Senjanovi\ifmmode~\acute{c}\else \'{c}\fi{}}}]{PhysRevD.58.115007}%
  \BibitemOpen
  \bibfield  {author} {\bibinfo {author} {\bibfnamefont {C.~S.}\ \bibnamefont
  {Aulakh}}, \bibinfo {author} {\bibfnamefont {A.}~\bibnamefont {Melfo}},
  \bibinfo {author} {\bibfnamefont {A.}~\bibnamefont {Ra\ifmmode~\check{s}\else
  \v{s}\fi{}in}}, \ and\ \bibinfo {author} {\bibfnamefont {G.}~\bibnamefont
  {Senjanovi\ifmmode~\acute{c}\else \'{c}\fi{}}},\ }\href {\doibase
  10.1103/PhysRevD.58.115007} {\bibfield  {journal} {\bibinfo  {journal} {Phys.
  Rev. D}\ }\textbf {\bibinfo {volume} {58}},\ \bibinfo {pages} {115007}
  (\bibinfo {year} {1998}{\natexlab{b}})}\BibitemShut {NoStop}%
\bibitem [{\citenamefont {Kuchimanchi}\ and\ \citenamefont
  {Mohapatra}(1993)}]{PhysRevD.48.4352}%
  \BibitemOpen
  \bibfield  {author} {\bibinfo {author} {\bibfnamefont {R.}~\bibnamefont
  {Kuchimanchi}}\ and\ \bibinfo {author} {\bibfnamefont {R.~N.}\ \bibnamefont
  {Mohapatra}},\ }\href {\doibase 10.1103/PhysRevD.48.4352} {\bibfield
  {journal} {\bibinfo  {journal} {Phys. Rev. D}\ }\textbf {\bibinfo {volume}
  {48}},\ \bibinfo {pages} {4352} (\bibinfo {year} {1993})}\BibitemShut
  {NoStop}%
\bibitem [{\citenamefont {Kuchimanchi}\ and\ \citenamefont
  {Mohapatra}(1995)}]{PhysRevLett.75.3989}%
  \BibitemOpen
  \bibfield  {author} {\bibinfo {author} {\bibfnamefont {R.}~\bibnamefont
  {Kuchimanchi}}\ and\ \bibinfo {author} {\bibfnamefont {R.~N.}\ \bibnamefont
  {Mohapatra}},\ }\href {\doibase 10.1103/PhysRevLett.75.3989} {\bibfield
  {journal} {\bibinfo  {journal} {Phys. Rev. Lett.}\ }\textbf {\bibinfo
  {volume} {75}},\ \bibinfo {pages} {3989} (\bibinfo {year}
  {1995})}\BibitemShut {NoStop}%
\bibitem [{\citenamefont {Barger}\ \emph {et~al.}(2009)\citenamefont {Barger},
  \citenamefont {Perez},\ and\ \citenamefont
  {Spinner}}]{PhysRevLett.102.181802}%
  \BibitemOpen
  \bibfield  {author} {\bibinfo {author} {\bibfnamefont {V.}~\bibnamefont
  {Barger}}, \bibinfo {author} {\bibfnamefont {P.}\ \bibnamefont {Fileviez~Perez}}, \
  and\ \bibinfo {author} {\bibfnamefont {S.}~\bibnamefont {Spinner}},\ }\href
  {\doibase 10.1103/PhysRevLett.102.181802} {\bibfield  {journal} {\bibinfo
  {journal} {Phys. Rev. Lett.}\ }\textbf {\bibinfo {volume} {102}},\ \bibinfo
  {pages} {181802} (\bibinfo {year} {2009})}\BibitemShut {NoStop}%
\bibitem [{\citenamefont {Holthausen}\ \emph {et~al.}(2010)\citenamefont
  {Holthausen}, \citenamefont {Lindner},\ and\ \citenamefont
  {Schmidt}}]{PhysRevD.82.055002}%
  \BibitemOpen
  \bibfield  {author} {\bibinfo {author} {\bibfnamefont {M.}~\bibnamefont
  {Holthausen}}, \bibinfo {author} {\bibfnamefont {M.}~\bibnamefont {Lindner}},
  \ and\ \bibinfo {author} {\bibfnamefont {M.~A.}\ \bibnamefont {Schmidt}},\
  }\href {\doibase 10.1103/PhysRevD.82.055002} {\bibfield  {journal} {\bibinfo
  {journal} {Phys. Rev. D}\ }\textbf {\bibinfo {volume} {82}},\ \bibinfo
  {pages} {055002} (\bibinfo {year} {2010})}\BibitemShut {NoStop}%
\bibitem [{\citenamefont {Ambroso}\ and\ \citenamefont
  {Ovrut}(2009)}]{1126-6708-2009-10-011}%
  \BibitemOpen
  \bibfield  {author} {\bibinfo {author} {\bibfnamefont {M.}~\bibnamefont
  {Ambroso}}\ and\ \bibinfo {author} {\bibfnamefont {B.~A.}\ \bibnamefont
  {Ovrut}},\ }\href {http://stacks.iop.org/1126-6708/2009/i=10/a=011}
  {\bibfield  {journal} {\bibinfo  {journal} {Journal of High Energy Physics}\
  }\textbf {\bibinfo {volume} {2009}},\ \bibinfo {pages} {011} (\bibinfo {year}
  {2009})}\BibitemShut {NoStop}%
\bibitem [{\citenamefont {Barger}\ \emph {et~al.}(2011)\citenamefont {Barger},
  \citenamefont {Fileviez~Perez},\ and\ \citenamefont
  {Spinner}}]{Barger:2010iv}%
  \BibitemOpen
  \bibfield  {author} {\bibinfo {author} {\bibfnamefont {V.}~\bibnamefont
  {Barger}}, \bibinfo {author} {\bibfnamefont {P.}~\bibnamefont
  {Fileviez~Perez}}, \ and\ \bibinfo {author} {\bibfnamefont {S.}~\bibnamefont
  {Spinner}},\ }\href {\doibase 10.1016/j.physletb.2011.01.015} {\bibfield
  {journal} {\bibinfo  {journal} {Phys. Lett.}\ }\textbf {\bibinfo {volume}
  {B696}},\ \bibinfo {pages} {509} (\bibinfo {year} {2011})},\ \Eprint
  {http://arxiv.org/abs/1010.4023} {arXiv:1010.4023 [hep-ph]} \BibitemShut
  {NoStop}%
\bibitem [{\citenamefont {Ghosh}\ \emph {et~al.}(2011)\citenamefont {Ghosh},
  \citenamefont {Senjanovic},\ and\ \citenamefont {Zhang}}]{Ghosh:2010hy}%
  \BibitemOpen
  \bibfield  {author} {\bibinfo {author} {\bibfnamefont {D.~K.}\ \bibnamefont
  {Ghosh}}, \bibinfo {author} {\bibfnamefont {G.}~\bibnamefont {Senjanovic}}, \
  and\ \bibinfo {author} {\bibfnamefont {Y.}~\bibnamefont {Zhang}},\ }\href
  {\doibase 10.1016/j.physletb.2011.03.039} {\bibfield  {journal} {\bibinfo
  {journal} {Phys. Lett.}\ }\textbf {\bibinfo {volume} {B698}},\ \bibinfo
  {pages} {420} (\bibinfo {year} {2011})},\ \Eprint
  {http://arxiv.org/abs/1010.3968} {arXiv:1010.3968 [hep-ph]} \BibitemShut
  {NoStop}%
\bibitem [{\citenamefont {Pérez}\ and\ \citenamefont
  {Spinner}(2009)}]{FileviezPerez2009251}%
  \BibitemOpen
  \bibfield  {author} {\bibinfo {author} {\bibfnamefont {P.}\ \bibnamefont
  {Fileviez~Perez}}\ and\ \bibinfo {author} {\bibfnamefont {S.}~\bibnamefont
  {Spinner}},\ }\href {\doibase
  http://dx.doi.org/10.1016/j.physletb.2009.02.047} {\bibfield  {journal}
  {\bibinfo  {journal} {Physics Letters B}\ }\textbf {\bibinfo {volume}
  {673}},\ \bibinfo {pages} {251 } (\bibinfo {year} {2009})}\BibitemShut
  {NoStop}%
\bibitem [{\citenamefont {Arason}\ \emph {et~al.}(1992)\citenamefont {Arason},
  \citenamefont {Casta\~no}, \citenamefont {Kesthelyi}, \citenamefont
  {Mikaelian}, \citenamefont {Piard}, \citenamefont {Ramond},\ and\
  \citenamefont {Wright}}]{PhysRevD.46.3945}%
  \BibitemOpen
  \bibfield  {author} {\bibinfo {author} {\bibfnamefont {H.}~\bibnamefont
  {Arason}}, \bibinfo {author} {\bibfnamefont {D.~J.}\ \bibnamefont
  {Casta\~no}}, \bibinfo {author} {\bibfnamefont {B.}~\bibnamefont
  {Kesthelyi}}, \bibinfo {author} {\bibfnamefont {S.}~\bibnamefont
  {Mikaelian}}, \bibinfo {author} {\bibfnamefont {E.~J.}\ \bibnamefont
  {Piard}}, \bibinfo {author} {\bibfnamefont {P.}~\bibnamefont {Ramond}}, \
  and\ \bibinfo {author} {\bibfnamefont {B.~D.}\ \bibnamefont {Wright}},\
  }\href {\doibase 10.1103/PhysRevD.46.3945} {\bibfield  {journal} {\bibinfo
  {journal} {Phys. Rev. D}\ }\textbf {\bibinfo {volume} {46}},\ \bibinfo
  {pages} {3945} (\bibinfo {year} {1992})}\BibitemShut {NoStop}%
\bibitem [{\citenamefont {Casta\~no}\ \emph {et~al.}(1994)\citenamefont
  {Casta\~no}, \citenamefont {Piard},\ and\ \citenamefont
  {Ramond}}]{PhysRevD.49.4882}%
  \BibitemOpen
  \bibfield  {author} {\bibinfo {author} {\bibfnamefont {D.~J.}\ \bibnamefont
  {Casta\~no}}, \bibinfo {author} {\bibfnamefont {E.~J.}\ \bibnamefont
  {Piard}}, \ and\ \bibinfo {author} {\bibfnamefont {P.}~\bibnamefont
  {Ramond}},\ }\href {\doibase 10.1103/PhysRevD.49.4882} {\bibfield  {journal}
  {\bibinfo  {journal} {Phys. Rev. D}\ }\textbf {\bibinfo {volume} {49}},\
  \bibinfo {pages} {4882} (\bibinfo {year} {1994})}\BibitemShut {NoStop}%
\bibitem [{\citenamefont {Electroweak}\ and\ \citenamefont
  {Groups}(2006)}]{Zprecision}%
  \BibitemOpen
  \bibfield  {author} {\bibinfo {author} {\bibfnamefont {S. Schael\ \emph {et~al.},}\ \bibnamefont {Electroweak}}\ and\ \bibinfo
  {author} {\bibfnamefont {H.~F.}\ \bibnamefont {Groups}},\ }\href {\doibase
  http://dx.doi.org/10.1016/j.physrep.2005.12.006} {\bibfield  {journal}
  {\bibinfo  {journal} {Physics Reports}\ }\textbf {\bibinfo {volume} {427}},\
  \bibinfo {pages} {257 } (\bibinfo {year} {2006})}\BibitemShut {NoStop}%
\bibitem [{\citenamefont {Takayama}\ and\ \citenamefont
  {Yamaguchi}(2000)}]{RefA}%
  \BibitemOpen
  \bibfield  {author} {\bibinfo {author} {\bibfnamefont {F.}~\bibnamefont
  {Takayama}}\ and\ \bibinfo {author} {\bibfnamefont {M.}~\bibnamefont
  {Yamaguchi}},\ }\href {\doibase 10.1016/S0370-2693(00)00726-7} {\bibfield
  {journal} {\bibinfo  {journal} {Phys. Lett.}\ }\textbf {\bibinfo {volume}
  {B485}},\ \bibinfo {pages} {388} (\bibinfo {year} {2000})},\ \Eprint
  {http://arxiv.org/abs/hep-ph/0005214} {arXiv:hep-ph/0005214 [hep-ph]}
  \BibitemShut {NoStop}%
\bibitem [{\citenamefont {Buchmuller}\ \emph {et~al.}(2007)\citenamefont
  {Buchmuller}, \citenamefont {Covi}, \citenamefont {Hamaguchi}, \citenamefont
  {Ibarra},\ and\ \citenamefont {Yanagida}}]{RefB}%
  \BibitemOpen
  \bibfield  {author} {\bibinfo {author} {\bibfnamefont {W.}~\bibnamefont
  {Buchmuller}}, \bibinfo {author} {\bibfnamefont {L.}~\bibnamefont {Covi}},
  \bibinfo {author} {\bibfnamefont {K.}~\bibnamefont {Hamaguchi}}, \bibinfo
  {author} {\bibfnamefont {A.}~\bibnamefont {Ibarra}}, \ and\ \bibinfo {author}
  {\bibfnamefont {T.}~\bibnamefont {Yanagida}},\ }\href {\doibase
  10.1088/1126-6708/2007/03/037} {\bibfield  {journal} {\bibinfo  {journal}
  {JHEP}\ }\textbf {\bibinfo {volume} {03}},\ \bibinfo {pages} {037} (\bibinfo
  {year} {2007})},\ \Eprint {http://arxiv.org/abs/hep-ph/0702184}
  {arXiv:hep-ph/0702184 [HEP-PH]} \BibitemShut {NoStop}%
\bibitem [{\citenamefont {Itoh}\ \emph {et~al.}(2006)\citenamefont {Itoh},
  \citenamefont {Okada},\ and\ \citenamefont {Yamashita}}]{IOY}%
  \BibitemOpen
  \bibfield  {author} {\bibinfo {author} {\bibfnamefont {H.}~\bibnamefont
  {Itoh}}, \bibinfo {author} {\bibfnamefont {N.}~\bibnamefont {Okada}}, \ and\
  \bibinfo {author} {\bibfnamefont {T.}~\bibnamefont {Yamashita}},\ }\href
  {\doibase 10.1103/PhysRevD.74.055005} {\bibfield  {journal} {\bibinfo
  {journal} {Phys. Rev.}\ }\textbf {\bibinfo {volume} {D74}},\ \bibinfo {pages}
  {055005} (\bibinfo {year} {2006})},\ \Eprint
  {http://arxiv.org/abs/hep-ph/0606156} {arXiv:hep-ph/0606156 [hep-ph]}
  \BibitemShut {NoStop}%
\end{thebibliography}%

\end{document}